\documentclass[twocolumn,showpacs]{revtex4} 
\usepackage{graphicx,color}

\newcommand{\etal}{{\it et al.}\ }

\begin{document}

\title{
First-principles structural optimization and electronic structure of\\
the superconductor picene for various potassium doping levels
}
\author{Taichi Kosugi$^{1}$}
\author{Takashi Miyake$^{1,2}$}
\author{Shoji Ishibashi$^1$}
\author{Ryotaro Arita$^{2,3,4}$}
\author{Hideo Aoki$^{5}$}

\affiliation{$^1$Nanosystem Research Institute ``RICS", AIST, Umezono, Tsukuba 305-8568, Japan}
\affiliation{$^2$Japan Science and Technology Agency (JST), CREST, Honcho, Kawaguchi, Saitama 332-0012, Japan}
\affiliation{$^3$Department of Applied Physics, University of Tokyo, Hongo, Tokyo 113-8656, Japan}
\affiliation{$^4$Japan Science and Technology Agency (JST), PRESTO, Kawaguchi, Saitama 332-0012, Japan}
\affiliation{$^5$Department of Physics, University of Tokyo, Hongo, Tokyo 113-0033, Japan}

\begin{abstract}
We theoretically explore the crystal structures of K$_x$picene, 
for which a new aromatic superconductivity has recently 
been discovered for $x=3$,  by 
systematically performing first-principles full structural optimization covering the concentration range $x=1$-$4$.  
The crystal symmetry (space group) 
of the pristine picene is shown to be preserved in all the optimized structures despite significant deformations of each picene molecule 
and vast rearrangements of herringbone array of molecules.  
For K$_x$picene ($x=1$-$4$) optimization indicates that (i) multiple 
structures exist in some cases, and (ii) dopants can enter  in the intralayer region as well as in the 
interlayer region between the stack of herringbone structures.  
In the electronic structure obtained with the local density approximation 
for the optimized structures,  the rigid-band approximation is invalidated 
for multifold reasons: 
the dopants affect the electronic properties 
not only through the rearrangement and distortion of molecules,
but also through hybridizations between the molecules and 
metal atoms.  As a consequence the resultant 
Fermi surface exhibits a variety of multiband 
structures which take diverse topology for K$_1$picene and K$_3$picene.
\end{abstract}

\pacs{74.20.Pq, 74.70.Kn, 74.70.Wz}

\maketitle

\section{Introduction}
The discovery of superconductivity in potassium-doped solid picene, 
K$_x$picene,  with $T_{\mathrm{c}} = 7$-$18$ K for a doping level $x \simeq 3$, recently reported by Mitsuhashi \etal\cite{bib:1330}, 
is seminal as the first {\it aromatic superconductor}.  
This is expected to open a new avenue in quest for organic superconductors.   The occurrence of superconductivity in the doped picene 
should not be an exceptional case, since the discovery has 
been followed by a more recent report of superconductivity in another of aromatic compounds, 
coronene, by Kubozono \etal\cite{bib:1996} with $T_{\mathrm{c}}$ up to $15$ K for K$_x$coronene with $x\simeq 3$, 
and also in doped phenanthrene with $T_{\mathrm{c}}$ up to $5$ K for $x\simeq 3$ by Wang \etal\cite{Wang}.  
These aromatic compounds are now forming a new class in the family of carbon-based superconductors that include graphite-intercalation compounds (GICs) and doped fullerenes.  
For GIC, KC$_8$, the first carbon-based superconductor with $T_{\mathrm{c}}= 0.1$ K\cite{bib:1114}, is followed by 
CaC$_6$ with the highest $T_{\mathrm{c}}= 11.6$ K among GICs\cite{bib:1108_10-11}. 
Doped fullerene K$_3$C$_{60}$, a superconductor with $T_{\mathrm{c}} = 20$ K \cite{bib:1104_1}, is 
followed by Cs$_2$RbC$_{60}$ with $T_{\mathrm{c}} = 33$ K\cite{bib:1104_4}, and Cs$_3$C$_{60}$ with $T_{\mathrm{c}} = 40$ K under $15$ kbar\cite{bib:1104_8} or 
$T_{\mathrm{c}} = 38$ K under $7$ kbar\cite{bib:1978}. 

The electronic structures of pristine and doped picene were first obtained by the 
present authors\cite{bib:1241} 
as an initial step to elucidate the mechanism of superconductivity.  
It has been revealed that the conduction band group is derived from the lowest unoccupied molecular orbital (LUMO) and LUMO+1 of a picene molecule. 
The conduction-band structure of solid picene, where the molecules are 
arrayed in a herringbone structure with two molecules per 
unit cell, is indeed shown to be accurately reproduced by a 
four-orbital tight-binding model downfolded in terms of 
the maximally-localized Wannier functions~\cite{bib:MLWF}.
We have also determined the dopant positions for K$_3$picene, 
where we have employed the experimental lattice parameters.
Andres \etal\cite{bib:1969}, on the other hand, 
provided two, fully-optimized geometries of K$_3$picene, one with a herringbone structure and the other with a laminar structure.  
Giovannetti and Capone\cite{bib:1971} performed an electronic structure calculations with different exchange-correlation functionals, 
and found that a hybrid functional leads to an antiferromagnetic state in K$_3$picene.  
Kim \etal\cite{bib:1970} demonstrated that an antiferromagnetic state is stabilized in K$_3$picene when the volume is increased.
They, as well as Giovannetti and Capone, suggested that a strong electron correlation exists in the system.  
This is reasonable in that the electronic correlation energy in solid picene is estimated to be $U \simeq 0.85$ eV with the momentum-dependent electron-energy-loss spectroscopy,\cite{bib:1997}
which is actually greater than the width of the LUMO- and LUMO+1-derived conduction bands.
This is also the case for potassium-doped picene, as will be demonstrated below.

Since the report of the superconductivity in potassium-doped picene,
a few theoretical studies aiming at clarifying its mechanism of superconductivity has been reported.  
These include phonon studies such as Kato \etal\cite{Kato1}, 
who have estimated the phonon frequencies and the electron-phonon coupling for various hydrocarbon molecules.  
For picene they found that the electron-phonon coupling is large enough to attain $T_{\mathrm{c}} \simeq 10$ K.  
Subedi and Boeri\cite{bib:1972} calculated the electron-phonon interactions in doped picene within the rigid-band approximation, and 
obtained coupling constants large enough to explain the experimental $T_{\mathrm{c}} \simeq 18$ K in K$_3$picene.
Casula \etal\cite{bib:2051} performed a more elaborate calculation, and obtained the electron-phonon coupling large enough to explain the experimental $T_{\mathrm{c}} \simeq 7$ K, 
to which they found the dopant and intermolecular phonon modes 
contribute significantly.  

However, one essential question is: what is so special about 
K$_3$picene as opposed to other values of $x$?  
The question is important, since it should be intimately related 
to another question: to what extent the aromatic superconductivity 
is general.  We can for instance remember that  the problem of 
why $A_3$fullerene ($A$: alkali metal elements) is special 
played an important role in elucidating its electronic properties.  
One experimentally unresolved situation 
is that, in the doped picene, both its 
atomic structure and even the number of dopant atoms actually 
accommodated in the samples have yet to be experimentally determined, 
although there are some estimates as discussed in Ref.\cite{bib:1996}.  
Given the circumstance, a theoretical prediction for 
the geometry of doped picene is thus desirable, especially 
for systematically varied doping levels.  This will also 
give a starting point to discuss whether the superconductivity is driven by a phonon mechanism, or by an electronic mechanism as 
in some classes of organic superconductors\cite{bib:1977}.  

This is exactly the purpose of the present work.  
For organic materials, not only the atomic positions but also the unit cell 
should be relaxed for a reliable optimization of the geometry.  
Hence we have performed 
here the full structural optimization of K$_x$picene,  covering the concentration range $x=1$-$4$.

\section{Method}

For the electronic structure calculation and the structural optimization in the present work, we adopt the density-functional theory 
with the projector augmented-wave (PAW) method~\cite{bib:PAW} 
with the Quantum MAterials Simulator (QMAS) package~\cite{bib:QMAS} 
in the local density approximation (LDA)\cite{bib:LDA}.  
The  Bloch pseudofunctions are expanded in plane waves, 
with an energy cutoff of $40$ Ry throughout the present work.
All the self-consistent field calculations are performed with $4 \times 4 \times 4$ $k$ points.  
The structural optimization allows the atomic coordinates and the cell parameters to be relaxed with no constraint imposed on them.
In each iteration of the self-consistent field calculation, the forces and stress\cite{bib:20} felt by the system are calculated.
The optimization is regarded as converged when the largest force and stress are smaller than
$5 \times 10^{-5}$ Ht/$a_{\mathrm{B}}$ and $5 \times 10^{-7}$ Ht/$a_{\mathrm{B}}^3$, respectively.
We employ the maximally-localized Wannier function technique\cite{bib:MLWF}, which has been recognized to be a standard method for 
describing electronic orbitals in solid.  
There, Wannier functions are constructed from the Bloch wave functions belonging to the specified electronic bands.
In the present work we extract the transfer integrals between the localized Wannier functions constructed for the solid picene.

\section{Results}

\begin{figure}[htbp]
\begin{center}
\includegraphics[keepaspectratio,width=6cm]{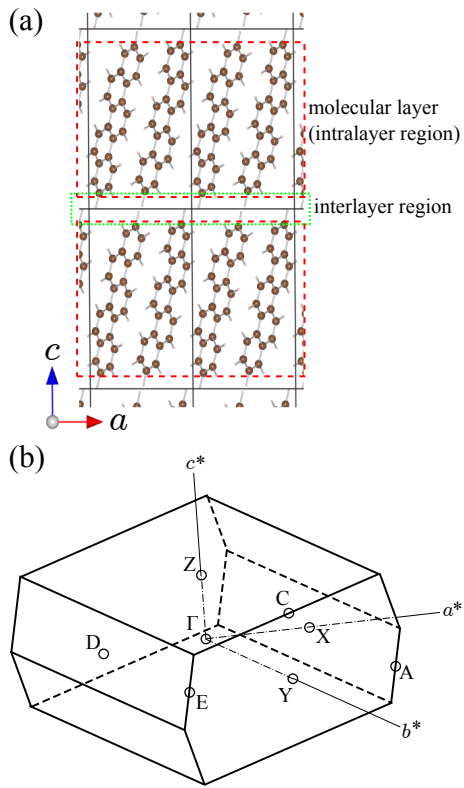}
\end{center}
\caption{
(Color online)
(a)
Crystal structure of pristine solid picene (drawn with VESTA\cite{bib:VESTA}), 
which comprises molecular layers stacked in the $c$ direction.
Solid lines delineate unit cells.
(b)
Typical first Brillouin zone of the doped picene optimized in the present work. 
Due to the crystal symmetry, we have $a^*\perp b^*$ and 
$b^*\perp c^*$, while $c^*$ and $a^*$ are not orthogonal to each other.
}
\label{Fig_pristine}
\end{figure}

Pristine solid picene comprises molecular layers stacked in the $c$ direction, as depicted in Fig. \ref{Fig_pristine}.
In each layer, picene molecules are in a herringbone arrangement, 
which means a unit cell contains two molecules. 
The crystal lattice is monoclinic with space group $P2_1$.
Its lattice parameters are reported to be $a=8.480, b=6.154, c=13.515$ 
in \AA, and the angle $\beta = 90.46^\circ$  between $a$ and $c$ axes from a single-crystal X-ray diffraction experiment.\cite{bib:1079} 
The samples used by Mitsuhashi \etal\cite{bib:1330} had the same lattice parameters.

Let us first classify the candidate positions into which dopants 
can be accommodated.  A conventional organic insulator, 
solid pentacene, where each molecule has the same 
chemical formula C$_{22}$H$_{14}$ as picene but five benzene rings 
are connected on a straight line with zigzag edges, is known to accommodate dopants in its interlayer region when doped with alkali elements,\cite{bib:alkali-doped_pentacene}
with its $c$ axis significantly expanded.
By contrast, the $c$ axis of picene is reported to 
shrink when potassium-doped  with $x=2.9$.\cite{bib:1330}
This suggests that the dopants have intruded into the intralayer region of solid picene.
To determine whether intralayer insertion of potassium atoms is energetically more favorable than interlayer insertion,
we performed in our previous work\cite{bib:1241} a structural optimization 
by locating a potassium atom on different positions on the $ab$ plane of the pristine picene as  initial geometries, 
and found that the intralayer insertion is in fact energetically favorable.
In the present work, we hence explore optimized geometries of the dopants 
with intralayer insertion.

\begin{figure*}[htbp]
\begin{center}
\includegraphics[keepaspectratio,height=11cm]{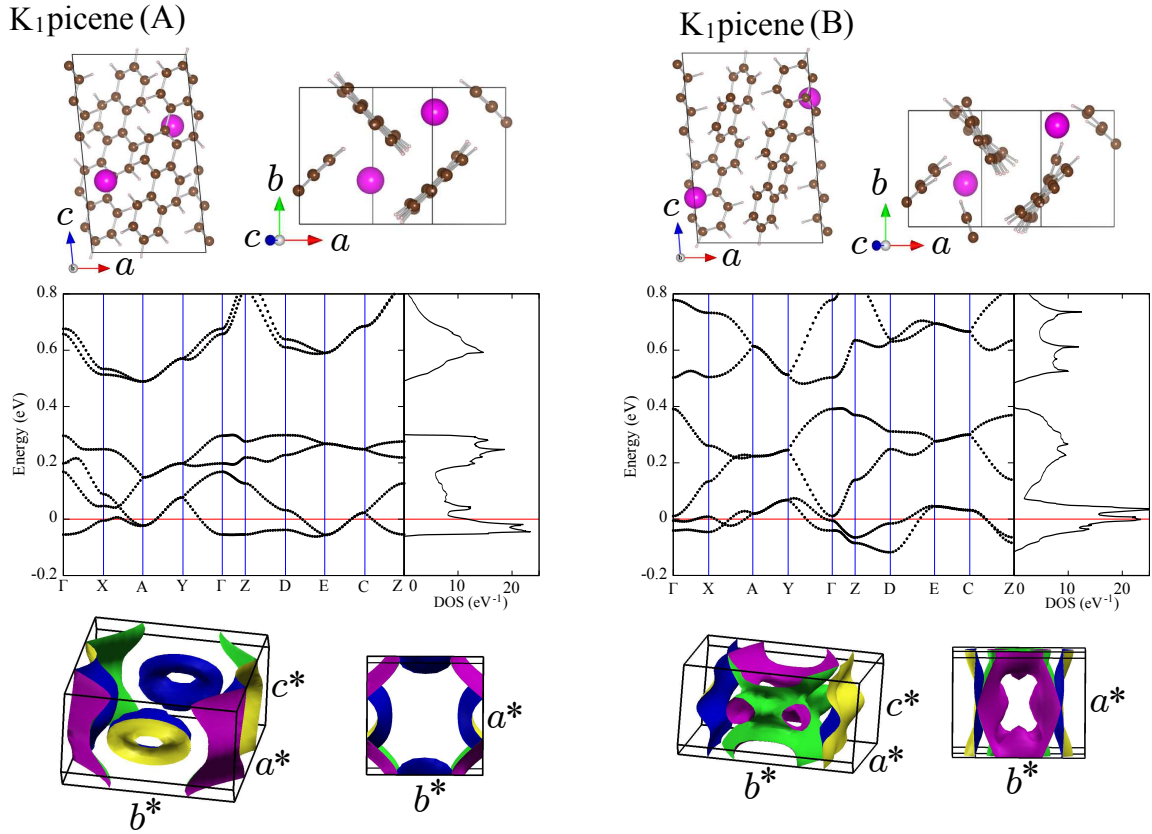}
\end{center}
\caption{
(Color online)
Geometries (large spheres: K atoms), electronic band structures, densities of states, and Fermi surfaces  of the K$_1$ structures.
The origins of energy are set to the respective Fermi level. 
}
\label{Fig_band_K1}
\end{figure*}

\begin{figure}[htbp]
\begin{center}
\includegraphics[keepaspectratio,height=8cm]{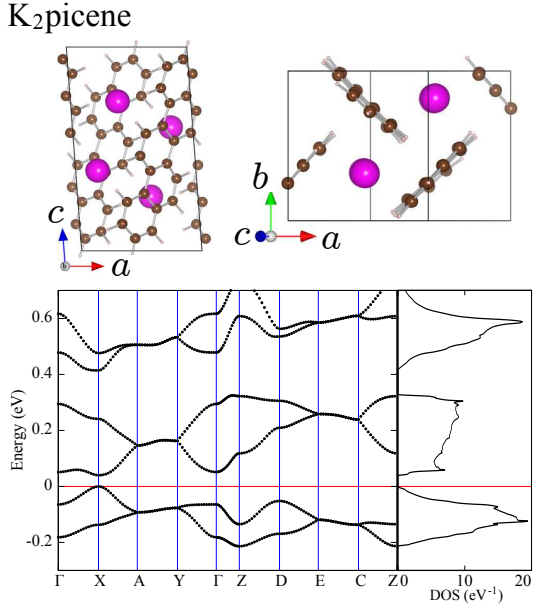}
\end{center}
\caption{
(Color online)
Geometry (large spheres: K atoms), electronic band structure, and 
density of states of the K$_2$picene.  
The origin of energy is set to the Fermi level.
}
\label{Fig_band_K2}
\end{figure}

\begin{figure}[htbp]
\begin{center}
\includegraphics[keepaspectratio,height=8cm]{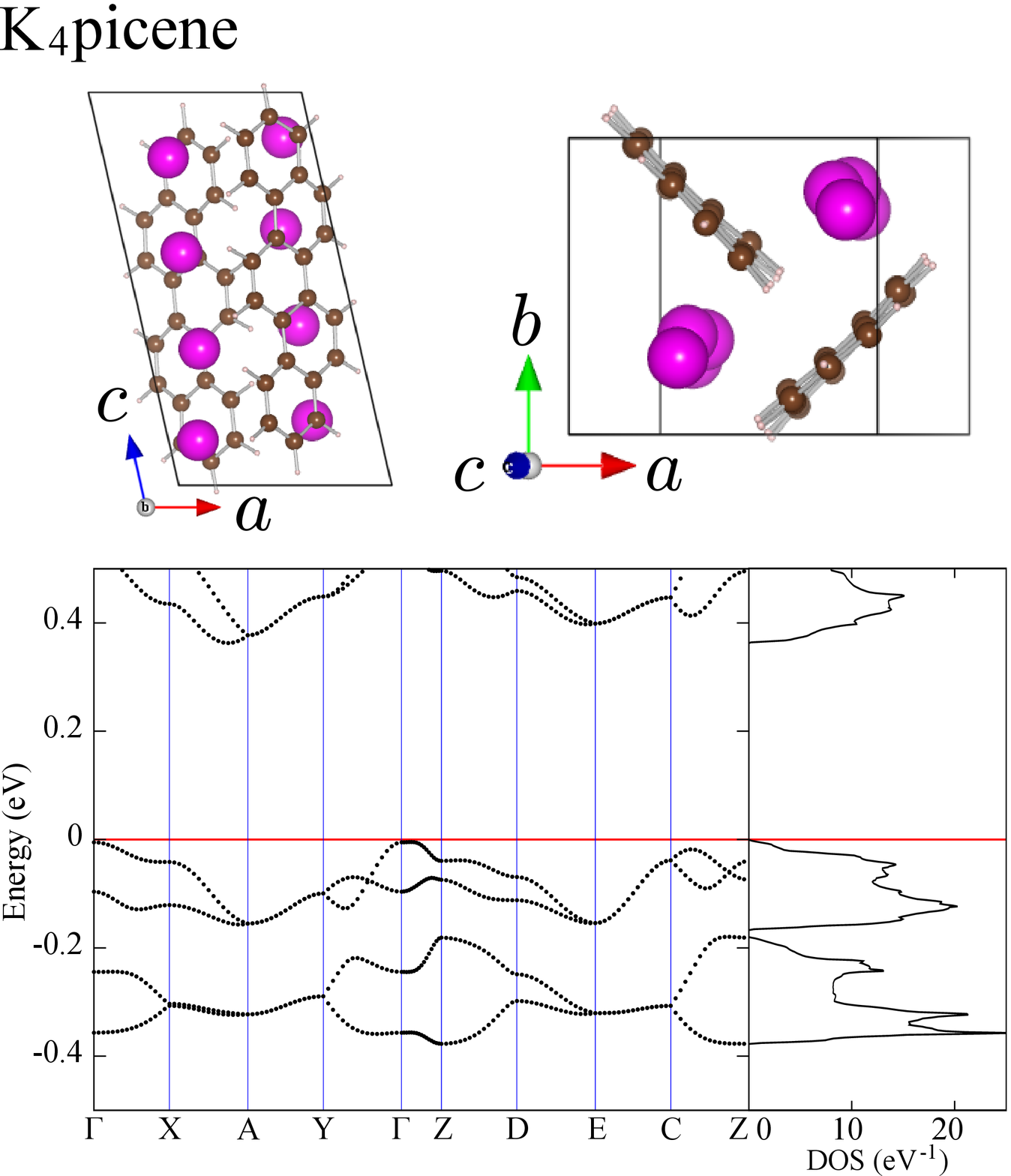}
\end{center}
\caption{
(Color online)
Geometry (large spheres: K atoms), electronic band structure, density of states, and Fermi surfaces of the K$_4$picene.
The origin of energy is set to the Fermi level.
}
\label{Fig_band_K4}
\end{figure}

The optimized lattice parameters and the internal coordinates of potassium atoms are summarized in Table \ref{table_latt}.
The crystal symmetry of pristine picene was found to be preserved in all the optimized doped structures, 
despite significant deformations of each molecule from the planar shape 
along with significant rearrangements of molecules upon doping.
The preserved symmetry contrasts with the situation in 
K$_3$coronene, for which we have shown that 
doping lowers the symmetry.\cite{kosugiPRB}
The geometries, the electronic band structures, the densities of states, and the Fermi surfaces for the optimized structures are shown 
for K$_1$picene in Fig. \ref{Fig_band_K1}, 
for K$_2$picene in Fig. \ref{Fig_band_K2}, 
for K$_3$picene in Fig. \ref{Fig_band_K3}, and 
for K$_4$picene in Fig. \ref{Fig_band_K4}.  
In each case, the conduction band group comprises 
four bands, which have main characters of LUMO and LUMO+1 
of an isolated picene molecule.  Since the herringbone structure 
has two molecules per unit cell, 
we have constructed four maximally-localized 
wave functions (WFs) from the four bands around the Fermi level of the individual structures.
For each of them, two WFs, $w_{\mathrm{h}}$ and $w_{\mathrm{l}}$, 
localized at each molecule with higher and lower orbital energies were obtained as shown in Fig. \ref{Fig_wanniers}.

\begin{table}[h]
\begin{center}
\caption{
Optimized lattice constants $a, b, c$ (in \AA), angle $\beta$ between $a$ and $c$ axes, volume $V$ of a unit cell, and internal coordinates of dopants in parentheses for K$_{x}$picene with $x = 1$-$4$.
}
\label{table_latt}
\begin{tabular}{lccccc}
\hline\hline
 & $a$ & $b$ & $c$ & $\beta$ & $V$ \\
\hline
K$_1$(A) & $7.267$ & $7.387$ & $12.698$ & $95.79$ & $688.876$ \\
 & \multicolumn{5}{c}{$(0.1820, 0.3202, 0.3615)$} \\
K$_1$(B) & $8.234$ & $6.554$ & $12.807$ & $94.65$ & $678.194$ \\
 & \multicolumn{5}{c}{$(0.0101, 0.3733, 0.2382)$} \\
K$_2$ & $7.237$ & $7.506$ & $12.624$ & $94.20$ & $683.868$ \\
 & \multicolumn{5}{c}{$(0.3916, 0.3155, 0.7333)$} \\
 & \multicolumn{5}{c}{$(0.1829, 0.3196, 0.3935)$} \\
K$_2$K$_1$ & $8.776$ & $6.394$ & $13.346$ & $94.03$ & $747.069$ \\
 & \multicolumn{5}{c}{$(0.3243, 0.2864, 0.6301)$} \\
 & \multicolumn{5}{c}{$(0.1806, 0.2848, 0.3225)$} \\
 & \multicolumn{5}{c}{$(0.1941, 0.2548, 0.0331)$} \\
K$_3$(A) & $7.421$ & $7.213$ & $14.028$ & $104.53$ & $726.848$ \\
 & \multicolumn{5}{c}{$(0.3357, 0.2790, 0.8241)$} \\
 & \multicolumn{5}{c}{$(0.2314, 0.2978, 0.5063)$} \\
 & \multicolumn{5}{c}{$(0.1167, 0.2792, 0.1818)$} \\
K$_3$(B) & $7.408$ & $7.223$ & $14.116$ & $105.93$ & $726.328$ \\
 & \multicolumn{5}{c}{$(0.4355, 0.2812, 0.8178)$} \\
 & \multicolumn{5}{c}{$(0.2247, 0.3007, 0.4932)$} \\
 & \multicolumn{5}{c}{$(0.0105, 0.2810, 0.1755)$} \\
K$_4$ & $7.511$ & $7.058$ & $14.230$ & $102.96$ & $735.230$ \\
 & \multicolumn{5}{c}{$(0.3039, 0.2585, 0.8344)$} \\
 & \multicolumn{5}{c}{$(0.2635, 0.3334, 0.5928)$} \\
 & \multicolumn{5}{c}{$(0.2441, 0.3202, 0.3477)$} \\
 & \multicolumn{5}{c}{$(0.1377, 0.2586, 0.1147)$} \\
\hline\hline
\end{tabular}
\end{center}
\end{table}

\subsection{K$_1$picene}

Let us start with K$_1$picene.  Two structures, 
K$_1$(A) and K$_1$(B), have been obtained.  
K$_1$(A) appears when we put one potassium atom per molecule in the intralayer 
region as the initial position of the optimization, 
while K$_1$(B) appears when we put the dopant in interlayer.  
In both structures, the dopants end up with intralayer positions after the optimization.  
The insertion of the dopants into the intralayer region supports the result of our previous work.\cite{bib:1241}   In 
K$_1$(B), with a total energy higher than in K$_1$(A)  by $0.316$ eV per molecule, the picene molecules are more strongly twisted than in K$_1$(A).  
Both systems have metallic electronic band structures.  
The width of the band group consisting of four  around their Fermi level 
is $0.36$ for K$_1$(A), and $0.52$ eV for K$_1$(B).

\subsection{K$_2$picene}

K$_2$picene structure was obtained by putting one potassium atom in the intralayer region and another in the interlayer region as the initial positions of the optimization, 
where we confirmed that an additional atom doped into K$_1$picene 
again prefers the intralayer region by 
moving to an intralayer position after the optimization.
The obtained K$_2$picene is an insulator, with a band gap of $0.038$ eV at X.
The widths of the highest occupied and the lowest unoccupied band groups are $0.21$ and $0.29$ eV, respectively.
By analyzing the projected density of states for the WFs, we find that each of the two band groups has comparable
weights for the two types of the Wannier functions $w_{\mathrm{h}}$ and $w_{\mathrm{l}}$.

\subsection{K$_3$picene}

\begin{figure*}[htbp]
\begin{center}
\includegraphics[keepaspectratio,height=19.5cm]{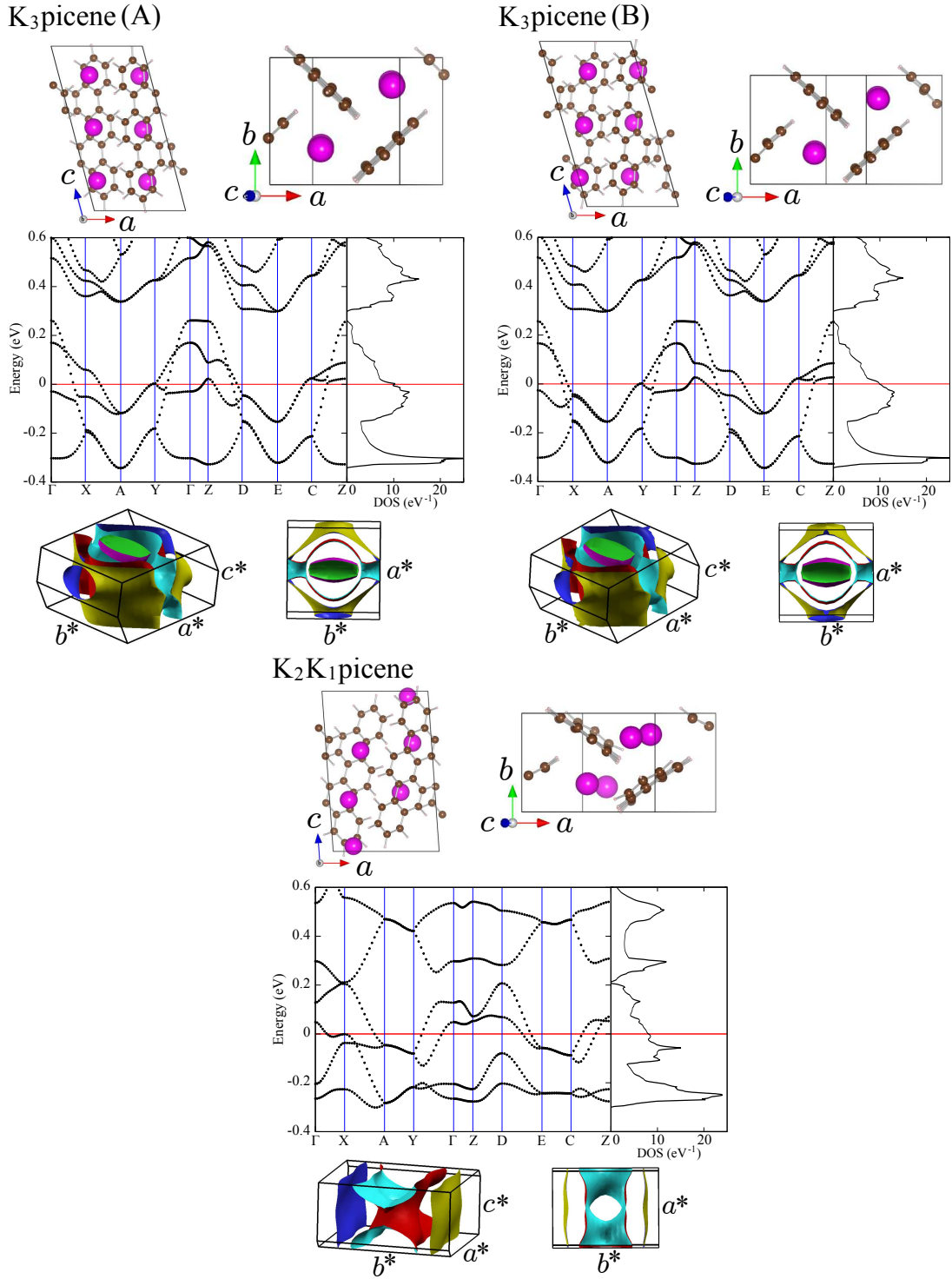}
\end{center}
\caption{
(Color online)
Geometries (large spheres: K atoms), electronic band structures, densities of states, and Fermi surfaces of the K$_3$picene.
The origins of energy are set to the respective Fermi level.
}
\label{Fig_band_K3}
\end{figure*}

For K$_3$picene 
we first obtaine K$_3$picene(A) structure 
by putting three potassium atoms in the intralayer region as the initial positions for the optimization.  
Andres \etal~\cite{bib:1969} obtained the monoclinic K$_3$picene structure with $a = 7.359, b = 7.361, c = 14.018$ in \AA \ and $\beta = 105.71^\circ$,
in which the potassium atoms are all in the intralayer region.
We have also tried a structural optimization starting from their lattice parameters and atomic coordinates.  
Then the geometry resulted in another structure, which we call K$_3$picene(B) here.  
Although the lattice parameters of K$_3$(A) and K$_3$(B) are close to each other with the difference in the total energy 
being only $1$ meV per molecule, they are distinct in the positions of dopants, as seen in Table \ref{table_latt} and in Fig. \ref{Fig_band_K3}.
K$_3$picene(B) has the lattice parameters and the dopant positions close to those of the K$_3$picene structure obtained by Andres \etal~\cite{bib:1969},
and we therefore consider the two structures to be the same.
K$_3$(A) and K$_3$(B) have metallic electronic band structures
with the same width $0.6$ eV of the band group consisting of four around their Fermi level.

Since the superconductivity in potassium-doped picene was observed for $x \simeq 3$,\cite{bib:1330, bib:1996}
more elaborate exploration of geometries is required than for other concentrations.
We have in fact found another structure, which we call K$_2$K$_1$picene,
by putting two potassium atoms in the intralayer region and one in the interlayer region as the initial positions for the optimization.
This structure is higher in the total energy than the K$_3$(A) structure by $0.465$ eV per molecule.  
However, the lattice parameters are closer to the experimental results\cite{bib:1330} of
$a = 8.707, b = 5.912, c = 12.97$ in \AA, and $\beta = 92.77^\circ$ for K$_{2.9}$picene.
The electronic band structure is again metallic for the K$_2$K$_1$ structure.
The four bands around the Fermi level, whose width is $0.51$ eV, are close to the higher bands 
as compared with those in the K$_3$(A) and K$_3$(B) structures.

Each of K$_3$(A) and K$_3$(B) has multiple Fermi surfaces 
comprising a pocket and complicated, connected surfaces.
The existence of a  Fermi pocket implies three-dimensional electron transfer 
integrals despite the layered structure.  
On the other hand, K$_2$K$_1$ has one-dimensional (a pair of 
planar) Fermi surfaces along with connected surfaces.  
We can also note that, while the Fermi surface of 
K$_3$coronene\cite{kosugiPRB} also consists of multiple surfaces, 
the topology of the Fermi surfaces for K$_3$coronene 
(consisting of 
a pair of quasi-one-dimensional, planar surfaces and 
another pocket) 
differs from those found here for 
K$_3$picene(A) and K$_3$picene(B) (consisting of multiple 
surfaces of two- or higher dimensional characters), 
and for K$_2$K$_1$picene (consisting of 
a pair of quasi-one-dimensional, planar surfaces and 
another, multiply-connected one).

\begin{figure}[htbp]
\begin{center}
\includegraphics[keepaspectratio,height=9cm]{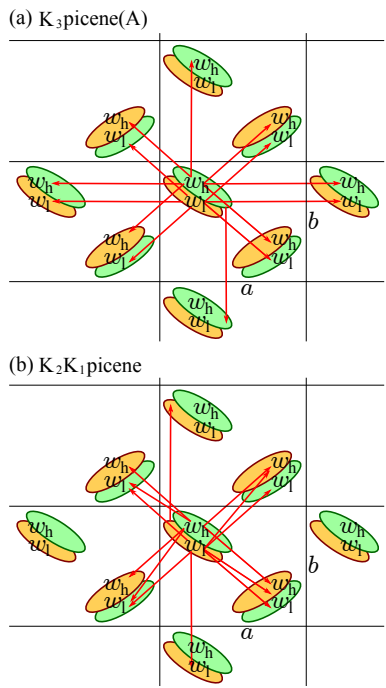}
\end{center}
\caption{
(Color online)
Significant ($|t| > 30$ meV) transfer integrals $t$ 
between Wannier functions, two at each molecule, 
are depicted with arrows 
for K$_3$picene(A) (a), and K$_2$K$_1$picene (b).
}
\label{Fig_transfers}
\end{figure}

The largest interlayer transfer integral of the WFs in K$_3$(A) (K$_2$K$_1$) is $13.6$ ($19.6$) meV,
while that in undoped picene is $14.6$ meV.
The larger interlayer transfer integrals in K$_2$K$_1$ than in K$_3$(A) originate in the presence of dopants in the interlayer region.
The widths of the bands around the Fermi level in these doped structures are larger than in undoped picene (0.39 eV\cite{bib:1241}),
though K$_3$(A) has the smaller interlayer transfer integrals.
It implies that their larger band widths come from the enhanced intralayer transfers of electrons due to the hybridization of the wave functions of potassium atoms and the molecular orbitals. 
In fact, the largest intralayer transfer integral in K$_3$(A) (K$_2$K$_1$) is $82.7$ ($70.6$) meV,
which is much larger than that in undoped picene, $57.9$ meV.
The intermolecular transfer integrals between the WFs of K$_3$(A)picene and K$_2$K$_1$picene are shown in Fig. \ref{Fig_transfers}.
It is seen that K$_3$(A) allows for the transfers between more distant WFs in the $a$ direction,
which leads to its more two-dimensional and wider band dispersion than K$_2$K$_1$.

\subsection{K$_4$picene}

We have obtained the K$_4$picene structure by putting three potassium atoms in the intralayer region and one in the interlayer region as the initial positions of the optimization, 
because the three dopants sitting on every other 
benzene rings out of five rings in each picene molecule may 
tend to push the fourth dopant to the interlayer region.  
In the optimized structure, all the four dopants end up with 
sitting in the intralayer region,
but two of them are rather close to the interlayer region due to the crowded space.
The system is an insulator with a band gap of $0.36$ eV.
The highest occupied band group consists of two bands, separated from the second highest group of two bands by $0.01$ eV.
The widths of the higher and lower band groups are $0.17$ and $0.20$ eV, respectively. 
By analyzing the projected density of states for the WFs, we found that the higher and lower occupied band groups consist mainly of $w_{\mathrm{h}}$ and $w_{\mathrm{l}}$, respectively.  
This contrasts with the situation in another insulator, K$_2$picene.

\subsection{Discussions}

For each concentration $x$, the lowest-energy structure shown above contains the dopants in the intralayer region.
This behavior is consistent with the observed tendency\cite{bib:1996} of the $c$ axis of K$_x$picene,
in which the $c$ axis shrinks monotonically with increasing $x$, implying that the dopants move into the intralayer region regardless of the concentration.
This is to be contrasted with Rb$_x$picene, whose $c$ axis was observed to expand up to $x = 2$ and shrink afterwards.

\begin{figure*}[htbp]
\begin{center}
\includegraphics[keepaspectratio,width=17cm]{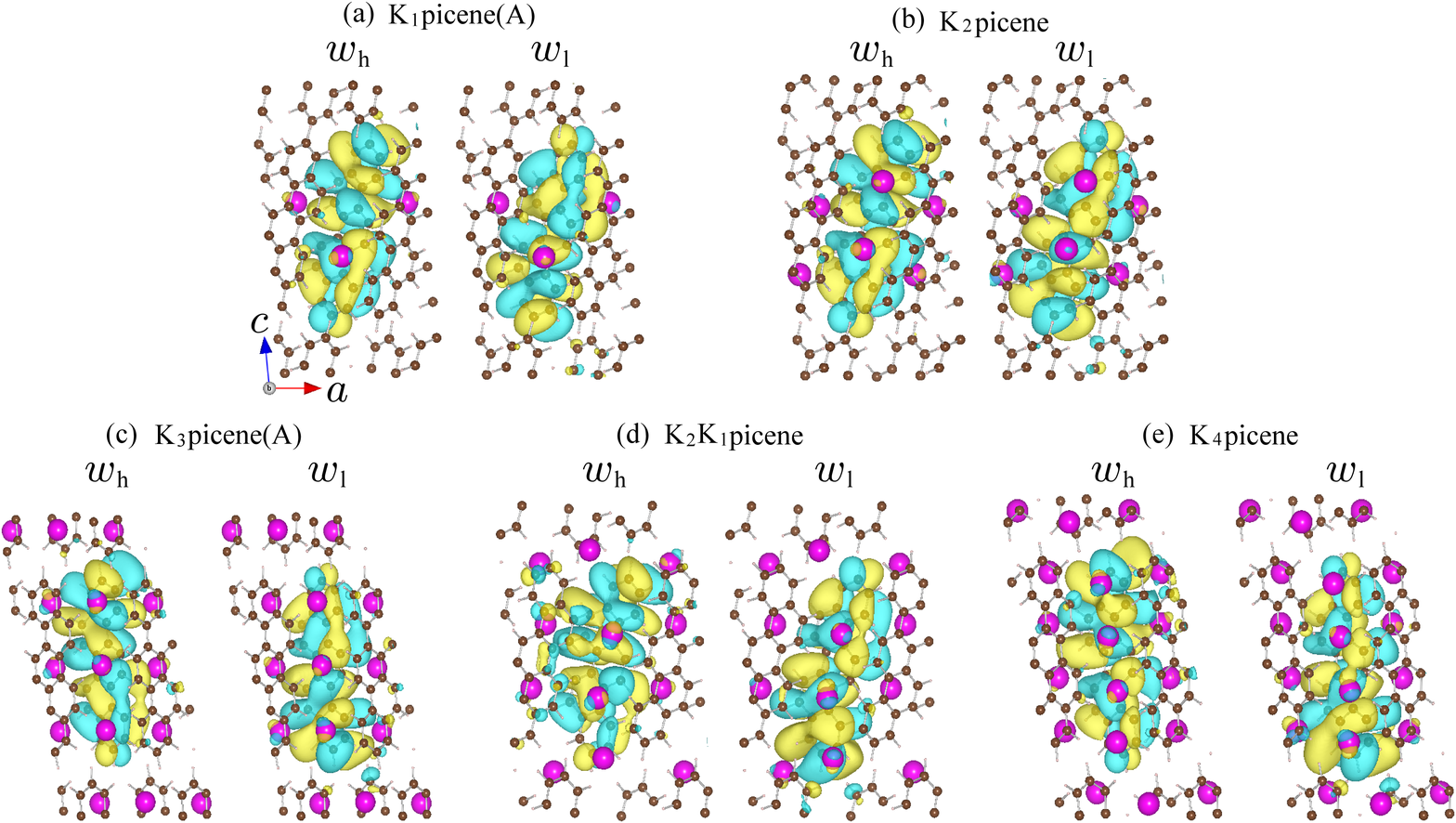}
\end{center}
\caption{
(Color online)
Maximally-localized Wannier functions in the optimized structures 
for K$_1$picene (a), K$_2$picene (b), K$_3$picene(A) (c), 
K$_2$K$_1$picene (d), and K$_4$picene (e).  
$w_{\mathrm{h}}$ and $w_{\mathrm{l}}$ stand for 
the higher- and lower-energy states, respectively.
}
\label{Fig_wanniers}
\end{figure*}

In Fig. \ref{Fig_wanniers} we have displayed the 
maximally-localized Wannier functions 
($w_{\mathrm{h}}$ and $w_{\mathrm{l}}$ with higher and lower orbital energies, respectively) in the optimized structures 
for all the cases of K$_1$picene to K$_4$picene.  
We can see that the WFs have basically the same features for all the structures in that all the WFs have LUMO and LUMO+1 orbital 
characters, although small but significant 
amplitudes found 
around the potassium atoms whose magnitudes depend on the 
doping level $x$. 
More precisely, we can show that 
$w_{\mathrm{h}}$ and $w_{\mathrm{l}}$ have characters 
of the linear combinations of LUMO and LUMO+1:
$w_{\mathrm{h,l}} = \phi_{\mathrm{LUMO}} \pm \phi_{\mathrm{LUMO + 1}}$, as in the undoped picene\cite{bib:1241}.

However, a key message in the present band structure results is 
that, although the orbital characters of the WFs are common to all the optimized geometries, their band structures very much differ.  
This is not surprising, since the molecular shape as well as the 
herringbone arrangement of molecules {\it distort} significantly 
with doping.  
More importantly, the lobes of molecular WFs extend outward the 
molecule via the potassium atoms, with 
significant {\it hybridizations} of WFs around the potassium atoms.
Specifically, we find that the small contributions from 
the potassium wave functions have a $p$ orbital character, 
as seen in Fig. \ref{Fig_wanniers}.

In other words, the rigid-band picture is invalidated 
for multifold reasons, i.e., through the rearrangement and distortion of molecules,
but also via 
hybridization of the potassium $p$ orbitals with the molecular orbitals of picene. 
Indeed, for all the optimized geometries we have also calculated their band structures (not shown)  with potassium atoms removed 
but the doping level is fixed by shifting the chemical potential
to find that the band structures are drastically changed from those of the doped systems.  
As a consequence the resultant 
Fermi surface exhibits a variety of multiband 
structures which take diverse topology for K$_1$picene and K$_3$picene.
This obviously provides an intriguing starting point for 
exploring mechanisms of superconductivity as a future work.

Final point of interest, about the molecular species dependence, 
is the following: we can note that 
the calculated band structures for all $x$ (including $x = 0$\cite{bib:1241}) have significant dispersion along the $c$ direction as well as along 
the $a$ and $b$ directions, which indicates isotropic components 
in the electron transfers.  
If we compare this with the situation in K$_x$coronene, 
we find that the band dispersion is more anisotropic in both undoped coronene and K$_3$coronene\cite{kosugiPRB}, which reflects 
the differences between the intermolecular distances in each direction.

\section{Summary}

We have systematically 
performed a full structural optimization for K$_x$picene, covering the 
doping level $x=1$-$4$.
The crystal symmetry of pristine picene of $P2_1$ was found to be preserved in all the optimized structures despite the deformation and 
rearrangements of the molecules upon doping.  
The optimized systems for K$_1$picene and K$_3$picene have metallic band structures,
while band gaps open around their Fermi levels for K$_2$picene 
and K$_4$picene.  
In each of the optimized systems, the electronic bands in the vicinity of the Fermi level are derived from LUMO and LUMO+1 of the aromatic 
molecule as in the undoped picene.
We found that the presence of dopants affects the electronic properties of all the doped structures.
The electronic structures obtained with the local density approximation 
for the optimized structures 
reveal that the rigid-band approximation is invalidated due to 
the rearrangement and deformation of molecules and the accompanied 
charge redistribution between the molecules and dopants.  
The resultant Fermi surface exhibits a variety of multiband 
structures which take diverse topology for K$_1$picene and K$_3$picene.

In all the optimized systems, the widths of the LUMO- and LUMO+1-derived bands do not exceed $0.6$ eV,
which are much smaller than the experimental electronic correlation energy $U \simeq 0.85$ eV,\cite{bib:1997} 
indicating that the material may be a strongly correlated electronic system.  
Further investigations for clarifying the material properties and the mechanism of superconductivity in doped picene should thus take into account the magnetic instability and the strong Coulomb interaction, as pointed out by Giovanetti and Capone\cite{bib:1971} and Kim \etal\cite{bib:1970}

Rb$_{3.1}$picene and Ca$_{1.5}$picene have also 
been reported to become superconducting with $T_{\mathrm{c}} \simeq 7$ K,\cite{bib:1996} 
which indicates that the transfer of three electrons from the alkali atoms to a picene molecule can lead to superconductivity in alkali-doped picene.
Theoretical exploration of their possible geometries is needed, 
where  Rb$_x$picene is expected 
to accommodate dopants in the interlayer region 
due to a larger atomic radius of Rb than in K$_x$picene.

\begin{acknowledgments} 
HA thanks Yoshihiro Kubozono for discussions.  
The present work is partially supported by the Next Generation Supercomputer Project,
Nanoscience Program from MEXT, Japan, and 
by Grants-in-aid No. 19051016 and 22104010 from MEXT, Japan and the JST PRESTO program.  
The calculations were performed with computational facilities at TACC, AIST as well as 
at Supercomputer Center of ISSP and at Information Technology Center, both at University of Tokyo.

\end{acknowledgments}

\end{document}